\documentclass[traditabstract]{aa}

\usepackage{graphicx}
\usepackage{hyperref}
\usepackage{amsmath}
\usepackage{txfonts}
\usepackage{natbib}
\usepackage{upgreek}
\newcommand{\dd}{\mathrm{d}}
\newcommand{\cc}{\mathrm{c}}
\newcommand{\vv}{\mathrm{v}}
\newcommand{\mm}{\mathrm{m}}
\renewcommand{\pi}{\uppi}
\newcommand{\ext}{{\mathrm{ext,}i}}

\graphicspath{{./figures/}}

\begin{document} 

\title{Triaxial collapse and virialisation of dark-matter haloes}

\author{C.~Angrick \and M.~Bartelmann}

\institute{Zentrum f\"ur Astronomie der Universit\"at Heidelberg, Institut f\"ur Theoretische Astrophysik, Albert-Ueberle-Str.~2, 69120 Heidelberg, Germany\\ \email{cangrick@ita.uni-heidelberg.de}}

\date{\emph{A\&A manuscript, version \today}}

\abstract{We reconsider the ellipsoidal-collapse model and extend it in two ways: We modify the treatment of the external gravitational shear field, introducing a hybrid model in between linear and non-linear evolution, and we introduce a virialisation criterion derived from the tensor virial theorem to replace the \textit{ad-hoc} criterion employed so far. We compute the collapse parameters $\delta_\cc$ and $\Delta_\vv$ and find that they increase with ellipticity $e$ and decrease with prolaticity $p$. We marginalise them over the appropriate distribution of $e$ and $p$ and show the marginalised results as functions of halo mass and virialisation redshift. While the hybrid model for the external shear gives results very similar to those obtained from the non-linear model, ellipsoidal collapse changes the collapse parameters typically by $(20\ldots50)\%$, in a way increasing with decreasing halo mass and decreasing virialisation redshift. We qualitatively confirm the dependence on mass and virialisation redshift of a fitting formula for $\delta_\cc$, but find noticeable quantitative differences in particular at low mass and high redshift. The derived mass function is in good agreement with mass functions recently proposed in the literature.}

\keywords{cosmology: theory -- methods: analytical -- cosmology: dark matter -- cosmology: cosmological parameters -- galaxies: clusters: general}

\maketitle

\section{Introduction}

The spherical-collapse model \citep[e.g.][]{Wang1998,Engineer2000,Mota2004,Bartelmann2006,Schaefer2008} is a fundamental ingredient in the theory of cosmic structure formation. Following the collapse of a slightly overdense, homogeneous sphere, it allows the derivation of two essential parameters; the overdensity $\Delta_\vv$ of a virialised halo compared to the mean or the critical cosmic density, and the critical linear density contrast $\delta_\cc$. The former is important because it allows relating sizes to masses of virialised structures, and the latter because it establishes a link between linear structure formation and the population statistics of collapsed haloes. Despite fundamental doubts as to the validity of such a simplified model for accurate cosmological predictions, the parameters derived from the spherical-collapse model or variants thereof allow surprisingly far-reaching predictions such as the halo mass function or the correlation properties of haloes, which are confirmed by numerical simulations.

The statistics of a Gaussian random field implies that spherical collapse should not occur. In fact, the probability distribution for the eigenvalues $\lambda_i$ of the Zel'dovich deformation tensor \citep{Doroshkevich1970} shows that spherical collapse has a vanishing probability,
\begin{equation}
\label{eq:Doroshkevich}
\begin{split}
p(\lambda_1,\lambda_2,\lambda_3)=&\frac{15^3}{8\pi\sqrt{5}\sigma^6}\exp\left(-\frac{3\delta^2}{\sigma^2}+\frac{15I}{2\sigma^2}\right) \\
&\times(\lambda_1-\lambda_2)(\lambda_2-\lambda_3)(\lambda_1-\lambda_3)
\end{split}
\end{equation}
because it requires $\lambda_1=\lambda_2=\lambda_3$. In this equation, $\delta\equiv\lambda_1+\lambda_2+\lambda_3$, $I\equiv\lambda_1\lambda_2+\lambda_2\lambda_3+\lambda_1\lambda_3$, and $\sigma^2$ denotes the variance of the matter power spectrum smoothed on a scale that corresponds to a halo mass $M$. Modifications of the original Press-Schechter mass function \citep{Press1974,Bond1991} motivated by ellipsoidal collapse \citep{Sheth2002} substantially improve the agreement of analytic predictions on the halo population with numerical simulations \citep[e.g. the \emph{Millennium simulation} by][]{Springel2005}.

Ellipsoidal collapse was analysed many times before \citep[see][for examples]{Bartelmann1993, Eisenstein1995, Bond1996}. Several authors have worked with the model by \citet{Bond1996}, generalising it for different cosmologies and introducing the scale factor $a$ as a time variable \citep{Monaco1995, Monaco1997, Monaco1998, Sheth2001, Sheth2002, Ohta2004}. We are reconsidering it here for two reasons.

First, we want to analyse how different assumptions on the treatment of the environment of a halo impact on the parameters $\Delta_\vv$ and $\delta_\cc$. Previous assumptions have been that the principal-axis system of the homogeneous ellipsoid is either identical with that of the external gravitational shear field, or that the two eigensystems do not coincide, introducing rotation and a deviation from the homogeneous mass profile. We introduce another assumption here, letting the eigensystems of the collapsing ellipsoid and its surrounding shear field follow each other until turn-around of the major principal axis and then decoupling both.

Second, we want to stop the collapse along any of the principal axes according to a physically motivated virialisation condition. Virialisation must be invoked to prevent the axes from collapsing to zero, and thus to be able to follow the entire collapse of an ellipsoid, i.e.~the collapse of its three principal axes. Conventionally, the collapse of each axis is stopped when $a_i(a)=0.177a$, where $a_i$ is the scale factor of the $i$-th axis and $a$ is the background scale factor \citep{Bond1996, Sheth2001}. In this way, $\Delta=a^3/(a_1a_2a_3)=178$ for spherical collapse in an Einstein-de Sitter universe at the time when the third axis virialises. However, the value $0.177$ has no fundamental physical motivation, and there is no guarantee for it not to be different for ellipsoidal rather than spherical collapse, or when cosmologies other than EdS are to be considered. We introduce a general virialisation condition based on the \emph{tensor virial theorem} that avoids introducting such an uncalibrated factor. We find substantial changes on both $\Delta_\vv$ and $\delta_\cc$ from both modifications and point out several discrepancies of our results with earlier studies.

We introduce the ellipsoidal-collapse model including our modifications in Sect.~\ref{sec:model}, present our results in Sect.~\ref{sec:results} and conclude in Sect.~\ref{sec:conclusions}.

\section{The model}
\label{sec:model}

In this Section, we shall briefly review the ellipsoidal-collapse model of \citet{Bond1996} for cosmologies with a cosmological constant introducing the scale factor $a$ as time variable. Furthermore, we shall present a physically motivated virialisation condition to stop the collapse of each axis, and show how to find the proper initial ellipticity and prolaticity as a function of mass and virialisation redshift.

\subsection{The evolution equations}
\label{subsec:evolution}

Let $a_i=R_i/R_\mathrm{pk}$ be the dimension-less principal axes of the ellipsoid, where $R_i$ with $1\leq i\leq3$ are its dimensional semi-major axes, and $R_\mathrm{pk}$ the size of a spherical top-hat corresponding to a mass $M=(4\pi/3) \rho_\mathrm{b} R_\mathrm{pk}^3$ with the cosmological background density $\rho_\mathrm{b}$. The evolution of the three principal axes $a_i$ with time $t$ in a cosmology with a cosmological constant $\Lambda=(8\pi G/c^2)\rho_\Lambda$ is given by
\begin{equation}
 \label{eq:basicEvolution}
\frac{\dd^2 a_i}{\dd t^2}=\frac{8}{3}\pi G \rho_\Lambda a_i-4\pi G \rho_\mathrm{b}a_i \left(\frac{1}{3}+\frac{\delta}{3}+\frac{b_i}{2}\delta+\lambda_{\ext}\right)
\end{equation}
\citep{Bond1996}, where $G$ is the gravitational constant and $c$ the speed of light. The density contrast of the ellipsoid with respect to the background density is $\delta=(\rho-\rho_\mathrm{b})/\rho_\mathrm{b}=a^3/(a_1a_2a_3)-1$. The parameters $b_i$ and $\lambda_{\ext}$ denote the internal and external contributions to the gravitational tidal shear which occur because of the deviation from sphericity. Generally, the total tidal field is described by the \emph{tidal field tensor} $\tens{T}$ with the elements $T_{ij}=\partial^2\Phi_\mathrm{P}/(\partial x_i\partial x_j)=T_{\mathrm{int,}ij}+T_{\mathrm{ext,}ij}$, where $\Phi_\mathrm{P}$ denotes the peculiar gravitational potential, and $T_{\mathrm{int,}ij}$ and $T_{\mathrm{ext,}ij}$ are the internal and external contributions to the shear, respectively. After a transformation into the ellipsoid's eigensystem, which is the same as the eigensystem of $\tens{T}$ in this model, the internal shear can be evaluated as
\begin{equation}
 \label{eq:defineIntShear}
b_i(t)= a_1(t)a_2(t)a_3(t)\int_0^\infty\frac{\dd\tau}{[a_i^2(t)+1]\prod_{k=1}^3[a_k^2(t)+1]^{1/2}}-\frac{2}{3}\,,
\end{equation}
while the external shear can be approximated by
\begin{equation}
 \label{eq:defineExtShear}
\lambda_{\ext}(t)\equiv\begin{cases}\dfrac{D_+(t)}{D_+(t_0)}\left[
                                  \lambda_i(t_0)-\dfrac{\delta(t_0)}{3}\right] &\text{(linear approx.)}\,, \\
                                  \dfrac{5}{4}b_i(t) &\text{(non-linear approx.)}\,,
                                 \end{cases}
\end{equation}
where $D_+$ is the linear growth factor, and the $\lambda_i$ are the eigenvalues of the Zel'dovich deformation tensor. See the first appendix of \citet{Bond1996} for details of the calculation. Here and in the following the index `0' refers to initial values. 

In linear approximation, the environment into which the ellipsoid is embedded evolves completely independently of it, whereas in the non-linear approximation it is tightly coupled to the ellipsoid. In Sect.~\ref{subsec:InfluenceExtShear} we shall introduce the \emph{hybrid model} as a third approximation for the external shear.

We can rewrite Eq.~\eqref{eq:basicEvolution} by using the scale factor $a$ as time variable using Friedmann's equation $\dot a^2=a^2H_0^2E^2(a)$, where $H_0$ is Hubble's constant. The expansion function of the universe, $E(a)$, introduces the dimension-less density parameters of matter and dark energy \emph{today}, $\Omega_\mm$ and $\Omega_\Lambda$, respectively. This gives
\begin{equation}
 \label{eq:basicEvolutionA}
\frac{\dd^2a_i}{\dd a^2}+\left[\frac{1}{a}+\frac{E'(a)}{E(a)}\right]\frac{\dd a_i}{\dd a}+\left[\frac{3\Omega_\mm}{2a^5 E^2(a)}C_i(a)-\frac{\Omega_\Lambda}{a^2 E^2(a)}\right]a_i=0\,,
\end{equation}
where a prime denotes differentiation with respect to $a$, and $C_i\equiv(1+\delta)/3+b_i/2+\lambda_{\ext}$. Equation~\eqref{eq:basicEvolutionA} defines a set of three coupled second-order differential equations, for which we need six independent initial conditions compatible with the Zel'dovich approximation for early times. These are provided by
\begin{align}
 \label{eq:initAxes}
a_i(a_0)&=a_0[1-\lambda_i(a_0)]\,, \\
\label{eq:initAxesVel}
\left.\frac{\dd a_i}{\dd a}\right|_{a_0}&=1-\lambda_i(a_0)-\left.\frac{\dd\ln D_+}{\dd\ln a}\right|_{a_0}\lambda_i(a_0)\approx 1-2\lambda_i(a_0)\,,
\end{align}
since $D_+(a)\approx a$ for $a\ll1$. Choosing $a_0=2\times10^{-5}$, this is comfortably fulfilled. In Appendix~\ref{ap:monaco}, we compare Eqs.~(\ref{eq:basicEvolutionA}--\ref{eq:initAxesVel}) with the results presented by \cite{Monaco1997} and see that they differ.

In the following, we shall assume that the eigenvalues $\lambda_i$ of the Zel'dovich tensor are ordered as $\lambda_1\geq\lambda_2\geq\lambda_3$, which implies that the ellipsoid first collapses along the direction~1, $a_1\rightarrow0$ first. At that time $\delta\rightarrow\infty$, and the collapse of the remaining two axes can no longer be followed so that we have to add a virialisation condition for each axis preventing their collapse to zero.

\subsection{The virialisation condition}
\label{subsec:virialisation}

Conventionally, the collapse of each axis is stopped when $a_i(a)=0.177a$, where $a_i$ is the scale factor of the $i$-th axis and $a$ the background scale factor \citep[e.g.][]{Bond1996,Sheth2001}. In this way, $\Delta=a^3/(a_1a_2a_3)=178$ for spherical collapse in an Einstein-de Sitter (EdS) universe at the time when the third axis is assumed to virialise. However, the value $0.177$ has no fundamental motivation in the physics of the collapse, and there is no guarantee that it remain unchanged in the case of ellipsoidal instead of spherical collapse, or in other cosmologies than Einstein-de Sitter.

We shall thus follow a different approach and present a physically well-motivated virialisation condition in the following to stop the collapse of each axis individually. We start from the \emph{tensor virial theorem} \citep[see][p.~213, p.~280]{Binney1987},
\begin{equation}
 \label{eq:tensorVirial}
\frac{1}{2}\frac{\dd^2I_{ij}}{\dd t^2}=2K_{ij}+\Pi_{ij}+W_{ij}+V_{ij}\,,
\end{equation}
where $\tens{I}$ is the \emph{moment of inertia tensor}, $\tens{K}$ and $\tens{\Pi}$ are the contributions to the \emph{kinetic energy tensor} coming from ordered and random motions, respectively, $\tens{W}$ is the \emph{potential energy tensor}, and $\tens{V}$ is the \emph{external potential energy tensor}. Their elements are generally defined as
\begin{equation}
\label{eq:defineTensors}
\begin{aligned}
 I_{ij}&\equiv\int_\mathcal{V}\dd^3x\,\rho x_i x_j\,, & W_{ij}&\equiv-\int_\mathcal{V}\dd^3x\,\rho x_i\frac{\partial\Phi}{\partial x_j}\,, \\
K_{ij}&\equiv\frac{1}{2}\int_\mathcal{V}\dd^3x\,\rho\overline{v}_i\overline{v}_j\,, & \Pi_{ij}&\equiv\int_\mathcal{V}\dd^3x\,\rho\sigma_{ij}^2\,,
\end{aligned}
\end{equation}
and
\begin{equation}
 \label{eq:defineExtTensor}
V_{ij}\equiv-\frac{1}{2}\int_\mathcal{V}\dd^3 x\,\rho\left(x_i\frac{\partial\Phi_\mathrm{ext}}{\partial x_j}+x_j\frac{\partial\Phi_\mathrm{ext}}{\partial x_i}\right)\,,
\end{equation}
where $\rho$ is the density of the fluid, $\Phi$ and $\Phi_\mathrm{ext}$ are the gravitational potentials of the ellipsoid itself and its surroundings, respectively, $\mathcal{V}$ is the volume which is integrated over, $\sigma_{ij}^2\equiv\overline{v_i v_j}-\overline{v}_i\overline{v}_j$ are the velocity dispersions, and the bar indicates averaging over $\mathcal{V}$.

We now specialise to the case of a homogeneous ellipsoid. For a stable mass configuration, the left-hand side of Eq.~\eqref{eq:tensorVirial} has to vanish for each component of the inertial tensor individually. Since the ellipsoid is assumed to be at rest and the ellipsoid's eigensystem is chosen as a reference frame, $\overline{v}_i=0$ so that $K_{ij}=0$ and $\Pi_{ij}=\int_\mathcal{V}\dd^3 x\,\rho \langle v_i^2\rangle \delta_{ij}$, with the Kronecker symbol $\delta_{ij}$. Note again that in this framework the eigensystems of the overdense ellipsoid and the gravitational tidal field are identical. For a homogeneous ellipsoid $v_i(x_i)=(\dot{a}_i/a_i)x_i$, thus
\begin{equation}
\label{eq:kinEnergy}
\Pi_{ij}=\frac{1}{5}\dot{a}_i^2 M \delta_{ij}\,,
\end{equation}
with the mass $M$ of the ellipsoid. The sum $W_{ij}+V_{ij}$ can be evaluated using $-\nabla(\Phi+\Phi_\mathrm{ext})=\ddot{\vec{x}}$ and Eq.~\eqref{eq:basicEvolution} to be
\begin{equation}
 \label{eq:potEnergy}
W_{ij}+V_{ij}=\int_\mathcal{V}\dd^3 x\, \rho \frac{\dd^2 a_i}{\dd t^2}\frac{x_i}{a_i}=\frac{1}{5}a_i^2 M \left(\frac{8\pi G}{3}\rho_\Lambda-4\pi G\rho_\mathrm{b}C_i\right)\delta_{ij}\,.
\end{equation}
Requiring that Eq.~\eqref{eq:tensorVirial} is fulfilled for each axis separately and introducing the scale factor $a$ as time variable yields the \emph{virialisation conditions} for the three axes $a_i$,
\begin{equation}
 \label{eq:virCondition}
\left(\frac{a_i'}{a_i}\right)^2=\frac{1}{a^2 E^2(a)}\left(\frac{3\Omega_\mm}{2a^3}C_i-\Omega_\Lambda\right)\,.
\end{equation}
When this condition is fulfilled for an axis together with $\dot{a}_i<0$, its collapse is stopped and its size is frozen in.

We emphasise that the former equation is consistent with the virialisation condition for spherical collapse in the EdS universe, where $R_\mathrm{v}/R_\mathrm{ta}=0.5$. The subscripts `v' and `ta' denote the time of virialisation and turn-around, respectively. For the special case of EdS, Eq.~\eqref{eq:virCondition} becomes
\begin{equation}
 \label{eq:virConditionSphere}
R=\frac{G M}{\dot{R}^2}\,,
\end{equation}
with $a_1=a_2=a_3\equiv R$. Using the parametric solution
\begin{equation}
\label{eq:paramSolution}
R=\frac{R_0}{2\delta_0}(1-\cos\theta)\,, \qquad t=\frac{3 t_0}{4\delta_0^{3/2}}(\theta-\sin\theta)
\end{equation}
\citep{Engineer2000} together with the relations $R_0/\delta_0=R_\mathrm{ta}$ and $R_0^3=9 GMt_0^2/2$ indeed gives the expected result that Eq.~\eqref{eq:virConditionSphere} is satisfied when $R=R_\mathrm{ta}/2$.

\begin{figure}[t]
\resizebox{\hsize}{!}{\includegraphics{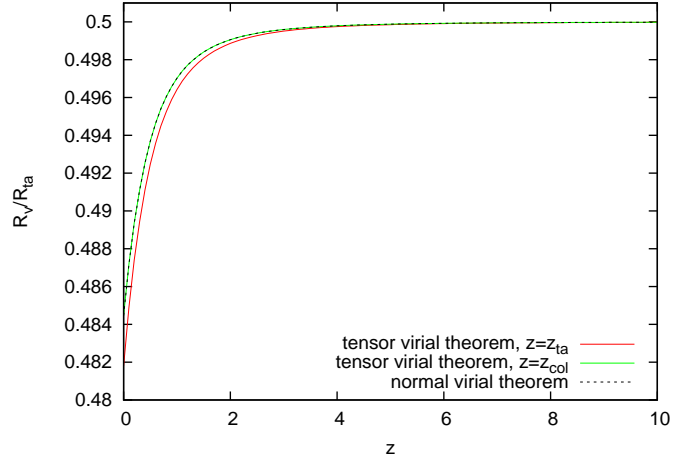}}
\caption{Comparison of the virialisation conditions from the scalar and tensor virial theorems. The curves differ slightly because they refer to different times. While the tensor virial theorem was evaluated at virialisation, the scaler virial theorem was evaluated at collapse. If we choose the collapse of the dark-matter halo as reference for $z$, both yield exactly the same result.}
\label{fig:virialisation}
\end{figure}

For bound objects in the $\Lambda$CDM model, the virialisation condition is
\begin{equation}
\label{eq:virialLCDM}
\langle E_\mathrm{kin}\rangle=-\frac{1}{2}\langle E_\mathrm{pot}\rangle+\langle E_\Lambda\rangle\,,
\end{equation}
where $\langle E_\mathrm{kin}\rangle$ is the average kinetic energy of the halo, $\langle E_\mathrm{pot}\rangle$ its average potential energy, and $\langle E_\Lambda\rangle$ the average effective potential energy contributed by the cosmological constant. Using $\langle E_\mathrm{pot}\rangle=-3GM^2/(5R)$ and $\langle E_\Lambda\rangle=-\Lambda MR^2/10$ for the homogeneous sphere as well as energy conservation between turn-around and virialisation, one arrives at a cubic equation for $R_\mathrm{v}/R_\mathrm{ta}$,
\begin{equation}
\label{eq:virialPolynomial}
2\eta\left(\frac{R_\mathrm{v}}{R_\mathrm{ta}}\right)^3-(2+\eta)\frac{R_\mathrm{v}}{R_\mathrm{ta}}+1=0\,,
\end{equation}
with $\eta=\Lambda R_\mathrm{v}^3/(3GM)$ \citep{Lahav1991}. The relevant solution of this equation agrees precisely with the condition derived from the tensor virial theorem, Eq.~\eqref{eq:virCondition}, as Fig.~\ref{fig:virialisation} shows. The small deviation occurs because we choose the time of virialisation rather than collapse as a reference ($z=z_\mathrm{v}$) when using the tensor virial theorem. If we use the collapse time ($z=z_\mathrm{col}$) instead, both conditions yield identical results.

\subsection{The initial ellipticity and prolaticity}
\label{subsec:epIni}

Equations~(\ref{eq:initAxes}, \ref{eq:initAxesVel}) imply that one has to choose initial values for the Zel'dovich deformation tensor to define the initial deviation of the principal axes and their time derivatives from the background. We shall show how they are chosen appropriately so that they comply with the assumed Gaussian nature of the Universe's initial conditions and represent a statistical average of haloes with the same mass $M$ and virialisation redshift $z_\vv$ but different shapes.

Starting from the probability distribution for the eigenvalues of the Zel'dovich deformation tensor, Eq.~\eqref{eq:Doroshkevich}, the conditional probability distribution for the \emph{ellipticity} $e\geq0$ and the \emph{prolaticity} $-e\leq p\leq e$, defined as
$e\equiv(\lambda_1-\lambda_3)/(2\delta)$ and $p\equiv(\lambda_1-2\lambda_2+\lambda_3)/(2\delta)$, respectively, was derived by \citet{Sheth2001} to be
\begin{equation}
 \label{eq:condProbEP}
g(e,p|\delta)=\frac{1125}{\sqrt{10\pi}}e(e^2-p^2)\left(\frac{\delta}{\sigma}\right)^5\exp\left[-\frac{5}{2}\left(\frac{\delta}{\sigma}\right)^2(3e^2+p^2)\right]\,.
\end{equation}
To compute a statistical average of any quantity $\xi(e,p)$ for a given halo that depends on $e$ and $p$, one should marginalise over the distribution $g(e,p)$,
\begin{equation}
 \label{eq:marginalise}
\langle \xi\rangle=\int_0^\infty\dd e\int_{-e}^e\dd p\, \xi(e,p) g(e,p)\,.
\end{equation}
However, a Taylor expansion of $\langle\xi\rangle$ up to second order in $e$ and $p$ around their mean values $\langle e\rangle$ and $\langle p \rangle$ under the distribution $g(e,p)$ gives
\begin{equation}
 \label{eq:Taylor}
\langle \xi\rangle=\xi\left(\langle e\rangle,\langle p\rangle\right)+\frac{1}{2}\left.\frac{\partial^2 \xi}{\partial e^2}\right|_{\langle e\rangle,\langle p\rangle} \sigma_e^2+\frac{1}{2}\left.\frac{\partial^2 \xi}{\partial p^2}\right|_{\langle e \rangle,\langle p\rangle} \sigma_p^2\;,
\end{equation}
where $\sigma_e^2$ and $\sigma_p^2$ are the variances for $e$ and $p$ according to the distribution $g$. Up to first order, $\langle\xi\rangle=\xi\left(\langle e\rangle,\langle p \rangle\right)$. Deviations occur only at second order. The expectation values for $e$ and $p$ as well as their variances are given by
\begin{align}
 \label{eq:expectation}
\langle e\rangle&=\frac{3\sigma}{\sqrt{10\pi}\delta}\,, & \langle p \rangle&=0\,, \\
\label{eq:variances}
\sigma_e^2&=\frac{(19\pi-54)\sigma^2}{60\pi\delta^2}\,, & \sigma_p^2&=\frac{\sigma^2}{20\delta^2}\;.
\end{align}
Since the variances are $\propto(\sigma/\delta)^2$, and $\sigma/\delta<1$, using the approximation $\langle\xi\rangle\approx\xi\left(\langle e\rangle,\langle p\rangle\right)$ introduces only a small error, which is $\sim$1\% for $\delta_\cc$ and $\sim$3\% for $\Delta_\vv$ for $\Lambda$CDM and OCDM. For EdS, the error is larger and $\sim$3\% for $\delta_\cc$ and $\sim$10\% for $\Delta_\vv$. But since the latter cosmology is scientifically only of low relevance and usually serves as a reference model only, the usage of the former approximation is well justified. Instead of sampling $\xi(e,p)$ at several points for $e$ and $p$, one only has to evaluate it once for $\langle e\rangle$ and $\langle p\rangle=0$.

Generally, the eigenvalues $\lambda_i$ are related to $e$ and $p$ by \citep[see e.g.][]{Bond1996,Bardeen1986}
\begin{align}
\label{eq:lambda1}
\lambda_1&=\frac{\delta}{3}(1+3e+p)=\frac{\delta}{3}+\frac{\sigma}{\sqrt{10\pi}}\,, \\
\label{eq:lambda2}
\lambda_2&=\frac{\delta}{3}(1-2p)=\frac{\delta}{3}\,, \\
\label{eq:lambda3}
\lambda_3&=\frac{\delta}{3}(1-3e+p)=\frac{\delta}{3}-\frac{\sigma}{\sqrt{10\pi}}\,,
\end{align}
where we have set $e=\langle e\rangle$ and $p=0$ in the last step.

\section{Results}
\label{sec:results}

In this Section, we show the results of the ellipsoidal-collapse model for the parameters $\delta_\cc$ and $\Delta_\vv$ for three different cosmologies and discuss how they are affected by the choice of the external-shear model.

\subsection{Influence of the external shear}
\label{subsec:InfluenceExtShear}

\begin{figure*}[t]
\centering
\includegraphics[width=0.33\textwidth]{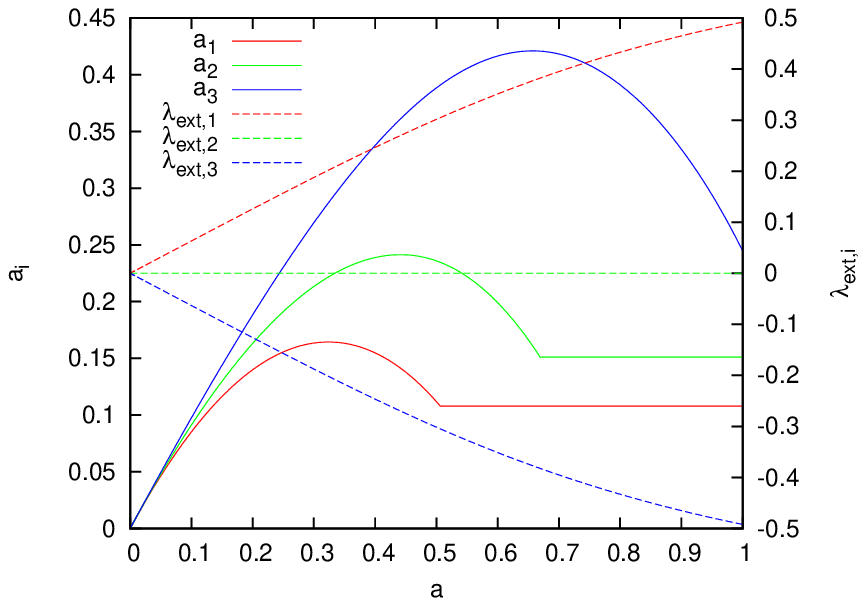}
\includegraphics[width=0.33\textwidth]{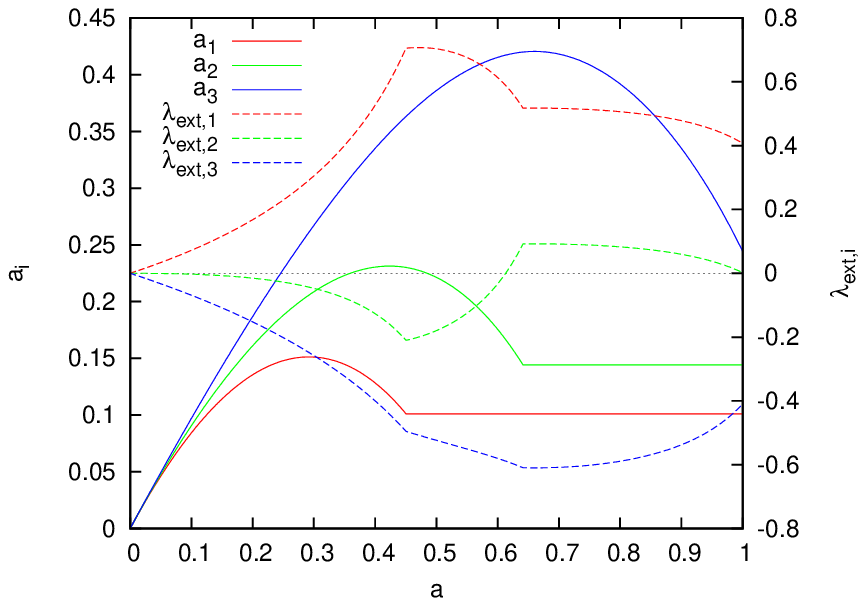}
\includegraphics[width=0.33\textwidth]{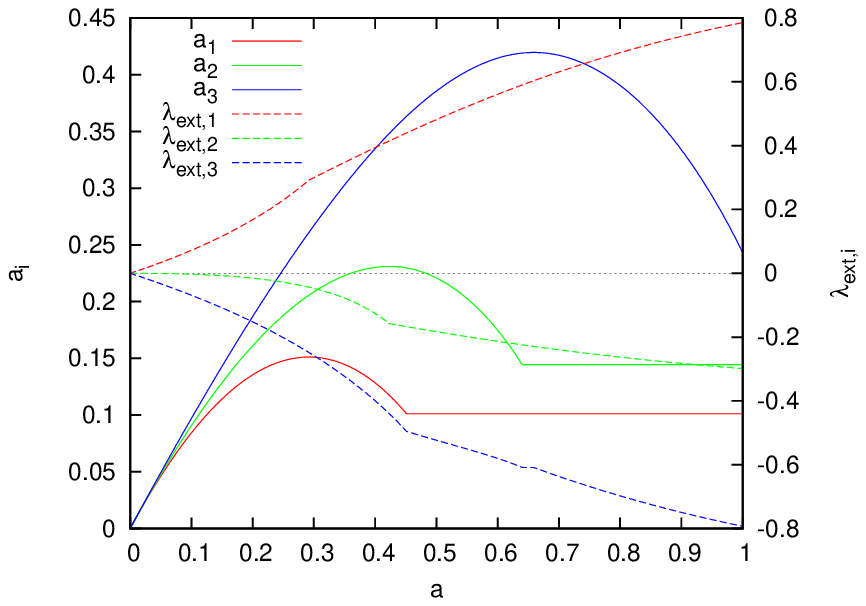}
\caption{Evolution of the principal axes and the external shear for a $10^{14}\ M_\odot/h$ dark-matter halo with $e=\langle e\rangle$ and $p=0$ in the reference $\Lambda$CDM model. \emph{Left panel:} Linear approximation. \emph{Central panel:} Non-linear approximation. \emph{Right panel:} Hybrid approximation.}
\label{fig:shearModels}
\end{figure*}

Figure~\ref{fig:shearModels} shows the evolution of the three principal axes and the eigenvalues of the external shear for a reference $\Lambda$CDM model with $\Omega_\mm=0.3$, $\Omega_\Lambda=0.7$, and $\sigma_8=0.8$ for three different models of the external shear and $e=\langle e\rangle$, $p=0$. The shear in the \emph{linear approximation} evolves completely smoothly over the entire collapse time. The evolution of each eigenvalue is given by $D_+(a)$ so that according to Eq.~\eqref{eq:defineExtShear}, $\lambda_\mathrm{ext,1}>0$, $\lambda_\mathrm{ext,2}=0$, and $\lambda_\mathrm{ext,3}=-\lambda_\mathrm{ext,1}<0$ at all times.

This is different in the \emph{non-linear approximation}: At early times, the evolution of the $\lambda_\ext$ is the same, but soon thereafter they start evolving non-linearly and $\lambda_\mathrm{ext,2}$ becomes slightly negative. Noticeably there are steps in the evolution of the external shear whenever an axis virialises because $\lambda_\ext\propto a_1a_2a_3$, and the evolution of this volume factor changes after virialisation of each axis. In the right panel, we introduce the \emph{hybrid approximation}: Initially, the evolution of $\lambda_\ext$ is described by the non-linear model. When one of the axes turns around, however, the corresponding eigenvalue of the external shear continues evolving linearly, i.e.~its value at turn-around is then scaled by $D_+(a)/D_+(a_\mathrm{ta})$.

We believe that the hybrid model is the preferred model for the evolution of the external shear since it takes into account that the evolution of the ellipsoid itself and its vicinity should be tightly coupled in the beginning. At turn-around, however, they are definitely decoupled so that choosing this moment to switch from non-linear to linear evolution seems appropriate. Hence, we will use the hybrid model for the evolution of the external shear in the following.

\begin{figure}[t]
\resizebox{\hsize}{!}{\includegraphics{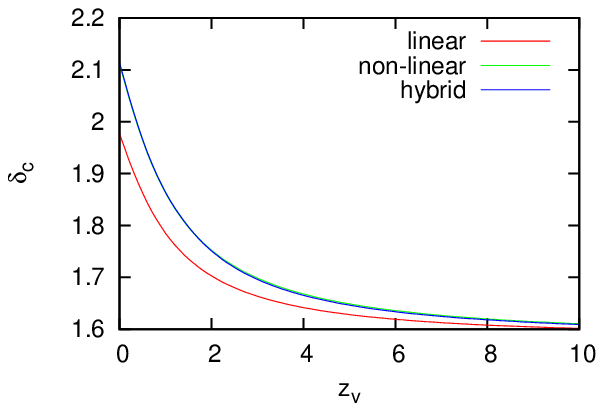} \includegraphics{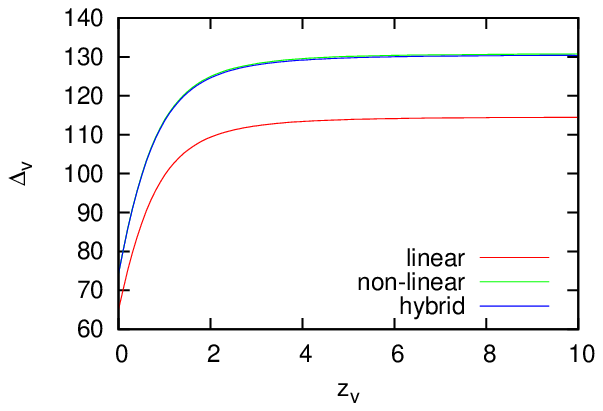}}\\
\resizebox{\hsize}{!}{\includegraphics{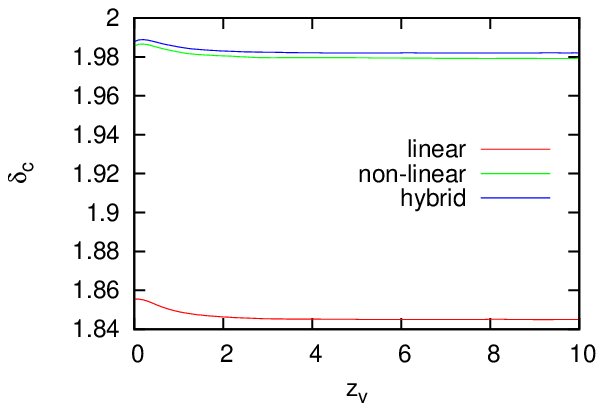} \includegraphics{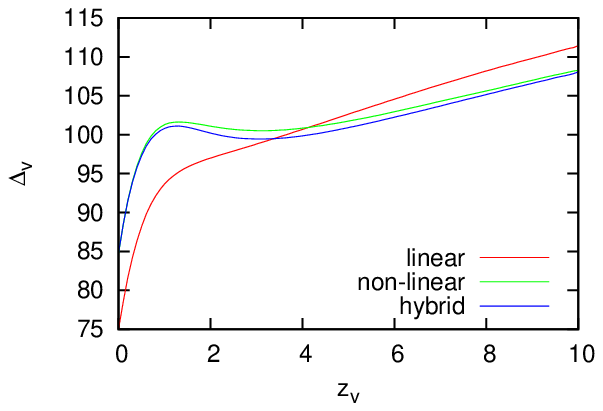}}
\caption{$\delta_\cc$ (\emph{left panels}) and $\Delta_\vv$ (\emph{right panels}) for three different models of the external shear. Both quantities were calculated for a $10^{14}\ M_\odot/h$ dark-matter halo in the reference $\Lambda$CDM model. \emph{Top panels:} $e=\langle e\rangle$ as a function of virialisation redshift $z_\vv$. \emph{Bottom panels:} $e=0.2$ for all $z_\vv$. In both cases, $p=0$.}
\label{fig:shearDeltas}
\end{figure}

The influence of the external-shear model on the parameters $\delta_\cc$ and $\Delta_\vv$ is shown in Fig.~\ref{fig:shearDeltas}. For a given mass and a given virialisation redshift $z_\vv$, the initial overdensity is chosen such that the third axis of the ellipsoid finally virialises at $z_\vv$. Both parameters can then be calculated by
\begin{equation}
 \label{eq:defineParam}
\delta_\cc=\frac{D_+(a_\vv)}{D_+(a_0)}\sum_{i=1}^3\lambda_i(a_0)\,,\qquad\Delta_\vv=\frac{\Omega_\mm(a_\vv) a_\vv^3}{a_1(a_\vv)a_2(a_\vv)a_3(a_\vv)}\,.
\end{equation}
In this case, $\Delta_\vv$ is defined with respect to the \emph{critical density} as it can be primarily found in the literature. For the definition with respect to the \emph{background density} one simply has to omit $\Omega_\mm(a_\vv)$. 

Using $\langle e\rangle$ as a function of virialisation redshift and $p=0$ (top panels of Fig.~\ref{fig:shearDeltas}), the dependence of $\delta_\cc$ and $\Delta_\vv$ on $z_\vv$ is almost the same for the non-linear and the hybrid models. However, both differ from the linear model, showing that the external shear is most important at the beginning of the ellipsoid's evolution. At that time, the non-linear and the hybrid models agree. While $\delta_\cc(z_\vv)$ is always smaller in the linear compared to the other two models, the curves for $\Delta_\vv(z_\vv)$ cross. This reflects the circumstance that the initial overdensity in the linear model is different from that in the two other models, leading to a different initial ellipticity and therefore to a completely different evolution history. This can be seen in the bottom panels of Fig.~\ref{fig:shearDeltas}, for which we have chosen $e=0.2$ independently of $z_\vv$. They also clearly show that a varying initial ellipticity drives primarily the evolution of $\delta_\cc$, whereas $\Delta_\vv$ also strongly varies for fixed $e$. In this case, both $\delta_\cc$ and $\Delta_\vv$ are smaller in the linear-shear model compared to the non-linear and the hybrid models.

\subsection{Parameters as function of mass and redshift}
\label{subsec:parameters}

Before we present general results for the parameters $\delta_\cc$ and $\Delta_\vv$, we should comment on a subtle but very important issue: Whenever we want to compare our results with the ordinary spherical-collapse model, we have to keep in mind that we calculate all quantities at the time when the third axis \emph{virialises}. Thus, we also have to compare these quantities those from the spherical-collapse model that are also calculated at the time of virialisation and not of collapse, i.e.~when $R=R_\mathrm{ta}/2$ and not $R=0$ for EdS. This leads to slightly lower reference values of $\delta_\cc$ and $\Delta_\vv$ since $z_\mathrm{col}<z_\vv$. Here the subscript `col' denotes collapse. Using the parametric solutions of \citet{Ohta2004} for the linear and the non-linear overdensity,
\begin{equation}
 \label{eq:parametric}
\delta_\mathrm{l}=\frac{3}{5}\left[\frac{3}{4}(\theta-\sin\theta)\right]^{2/3},\qquad \Delta_\mathrm{nl}=\frac{9}{2}\frac{(\theta-\sin\theta)^2}{(1-\cos\theta)^3}\,,
\end{equation}
respectively, and $\theta=3\pi/2$ at virialisation, we find that $\delta_\cc=1.583$ and $\Delta_\vv=147$ for the EdS universe at $R=R_\vv$ independent of $z_\vv$. Recently, \citet{Lee2009} arrived at the same values when accounting for the time of virialisation instead of collapse.

\begin{figure}[t]
\resizebox{\hsize}{!}{\includegraphics{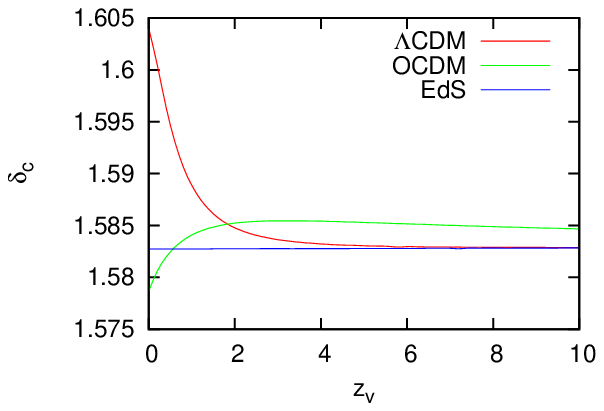} \includegraphics{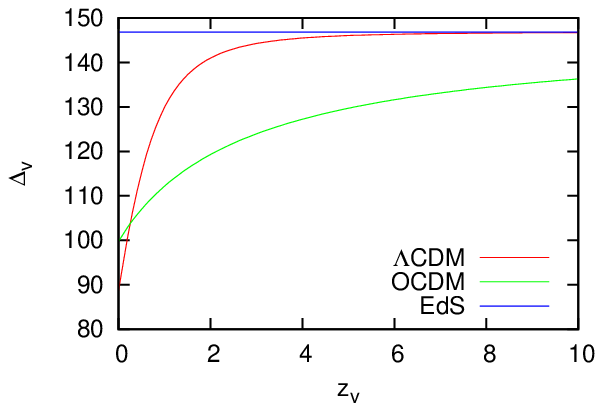}}\\
\resizebox{\hsize}{!}{\includegraphics{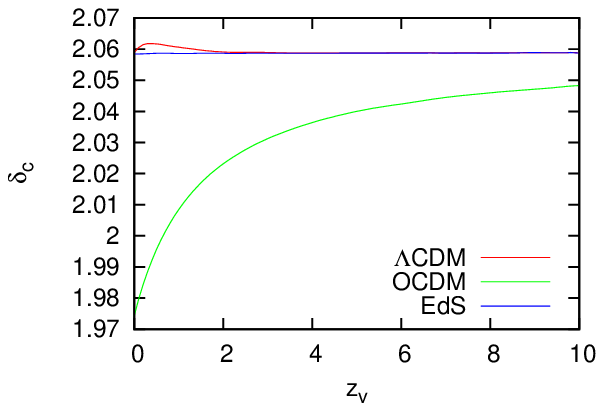} \includegraphics{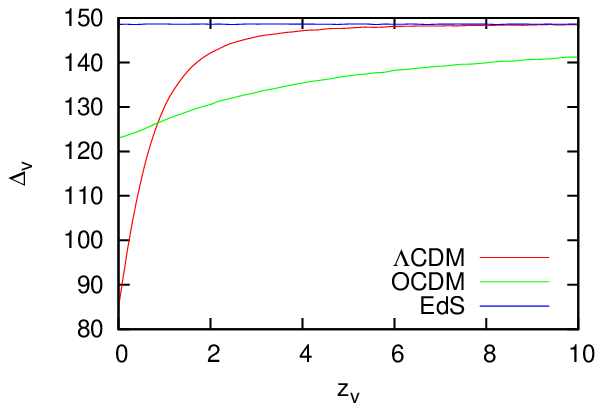}}
\caption{$\delta_\cc$ (\emph{left panels}) and $\Delta_\vv$ (\emph{right panels}) for three different cosmological models and fixed initial $e$ and $p$. \emph{Top panels:} Spherical case ($e=p=0$). \emph{Bottom panels:} Triaxial case with $e=0.2$ and $p=-0.1$.}
\label{fig:deltasFixed}
\end{figure}

The top panels of Fig.~\ref{fig:deltasFixed} show $\delta_\cc$ and $\Delta_\vv$ for three different cosmologies for $e=p=0$, i.e.~spherical systems. The OCDM cosmology is the same as our reference $\Lambda$CDM model except that $\Omega_\Lambda=0$. For the EdS model we set $\Omega_\mm=1$ and $\Omega_\Lambda=0$. Indeed, for the EdS universe the constant values derived analytically are also reproduced by solving Eq.~\eqref{eq:basicEvolutionA} numerically. This demonstrates again that Eqs.~(\ref{eq:basicEvolutionA}, \ref{eq:virCondition}) are fully consistent with the well-known spherical-collapse model. Note that the qualitative behaviour of $\Delta_\vv(z_\vv)$ is the same as $\Delta_\vv(z_\mathrm{col})$ \citep[compare e.g. with][]{Bartelmann2006}. However, there is a difference for the critical linear overdensity whose shape as a function of $z_\vv$ differs substantially from the shape as a function of $z_\mathrm{col}$. This should illustrate that the time chosen in the model when virialisation actually occurs ($z_\vv$ or $z_\mathrm{col}$) can already have substantial impact on the qualitative behaviour of relevant quantities as a function of redshift, hence it is not necessarily a consequence of ellipsoidal collapse alone.

The bottom panels of Fig.~\ref{fig:deltasFixed} show $\delta_\cc$ and $\Delta_\vv$ for a triaxial halo with $e=0.2$ and $p=-0.1$. Also in this case both parameters are independent of $z_\vv$ for the EdS universe, although $\delta_\cc$ changes from $1.583$ to $2.058$, while $\Delta_\vv$ stays almost the same, $148$ instead of $147$. Interestingly, the most drastic changes in the shapes of both parameters occur for the OCDM model, for which the total density is only approximately a third of the critical density, while for the $\Lambda$CDM model their changes are small.

\begin{figure}[t]
\resizebox{\hsize}{!}{\includegraphics{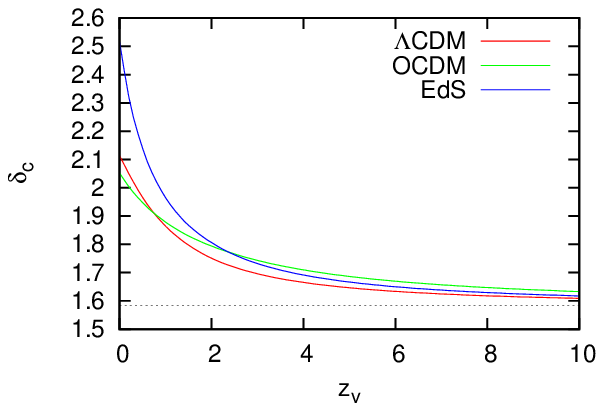} \includegraphics{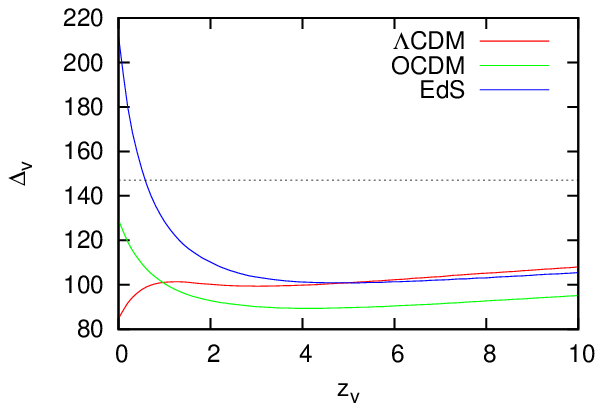}}\\
\resizebox{\hsize}{!}{\includegraphics{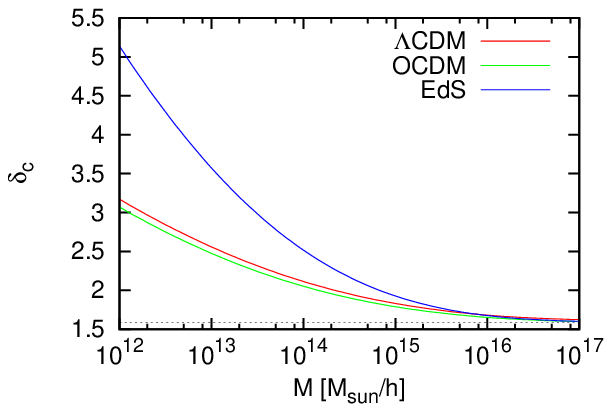} \includegraphics{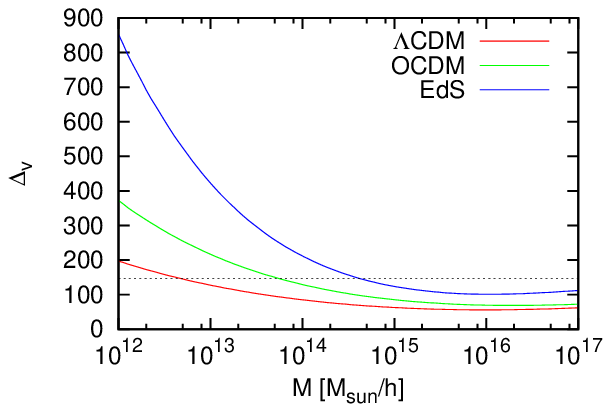}}
\caption{$\delta_\cc$ (\emph{left panels}) and $\Delta_\vv$ (\emph{right panels}) as a function of halo mass and virialisation redshift for three different cosmologies marginalised over $e$ and $p$. \emph{Top panels:} Dependence on virialisation redshift $z_\vv$ for a halo with $M=10^{14}\ M_\odot/h$. \emph{Bottom panels:} Dependence on mass $M$ for a halo with $z_\vv=0$. Thin dashed lines show the reference values from spherical collapse in the EdS universe, $\delta_\cc=1.583$ and $\Delta_\vv=147$.}
\label{fig:deltasMZ}
\end{figure}

The influence of the halo mass $M$ and the virialisation redshift $z_\vv$ on both $\delta_\cc$ and $\Delta_\vv$ are illustrated in Fig.~\ref{fig:deltasMZ}. For both decreasing mass and decreasing redshift, $\delta_\cc$ is a monotonically decreasing function, approaching the reference values from the spherical-collapse model in the EdS universe for large masses and high virialisation redshifts, the situation for $\Delta_\vv$ is much more complicated. For all three models, it has a minimum at redshifts $4$--$5$ and at a mass of $\sim10^{16}\ M_\odot/h$. For smaller values of mass and redshift, it is a monotonically decreasing function of both $M$ and $z_\vv$ for the OCDM and EdS models. It is also monotonically decreasing as a function of $M$ in the $\Lambda$CDM model, but reaches a maximum at $z_\vv\sim 1$. This is a direct result of the definition of $\Delta_\vv$ with respect to the \emph{critical density}. If it was defined with respect to the \emph{background density}, the factor $\Omega_\mm(a_\vv)$ would not appear in Eq.~\eqref{eq:defineParam}, and all three curves would increase with decreasing $z_\vv$.

In Fig.~\ref{fig:deltasMZ} one can clearly see that the intervals that are covered for both $\delta_\cc$ and $\Delta_\vv$ are the largest for the EdS model, indicating that there is a stronger dependence of the ellipsoid's evolution on the total amount of matter in the Universe compared to the size of the cosmological constant if varying initial ellipticities are taken into account.

For either $M\rightarrow\infty$ or $z_\vv\rightarrow\infty$, both $\delta_\cc$ and $\Delta_\vv$ must reach the reference values for spherical collapse in the EdS universe since the initial ellipticity $\langle e\rangle\propto\sigma/\delta$ and $\sigma$ both decrease with increasing mass, and $\delta$ has to be higher the earlier the structure is required to collapse. This expected behaviour can be clearly seen for $\delta_\cc$, whereas for $\Delta_\vv$ this happens finally for \emph{very} large $M$ and $z_\vv$. We should stress again in this context that a crucial portion of the dependence on mass and virialisation redshift is driven by the change in the initial ellipticity, comparing Figs.~\ref{fig:deltasFixed} and \ref{fig:deltasMZ}.

\begin{figure}[t]
\resizebox{\hsize}{!}{\includegraphics{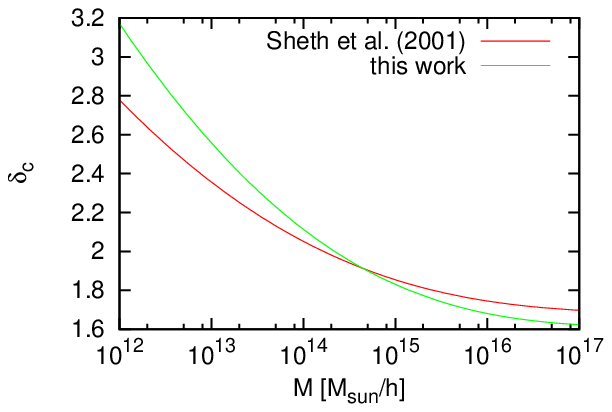} \includegraphics{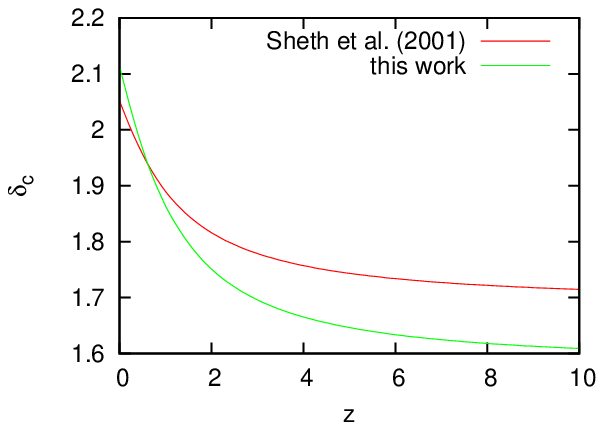}}
\caption{Comparison of the fitting formula of \citet{Sheth2001} for $\delta_\cc(M,z)$ with the results of this work for the reference $\Lambda$CDM model. \emph{Left panel:} $\delta_\cc$ as a function of mass and $z=0$. \emph{Right panel:} $\delta_\cc$ as a function of redshift and $M=10^{14}\ M_\odot/h$.}
\label{fig:Sheth}
\end{figure}

We compare in Fig.~\ref{fig:Sheth} the results of this work for $\delta_\cc$ as a function of mass and redshift with the fitting formula of \citet{Sheth2001} given by
\begin{equation}
\label{eq:fitting}
\delta_\cc(M,z)=\delta_\mathrm{c,sph}(z)\left\{1+0.47\left[\frac{\sigma^2(M,z)}{\delta_\mathrm{c,sph}^2(z)}\right]^{0.615}\right\}\,,
\end{equation}
where $\delta_\mathrm{c,sph}(z)$ is the redshift-dependent linear overdensity of the usual spherical-collapse model. The left panel shows that the dependence on mass is similar for both the fitting formula and the result of this work. However, the fit by \citet{Sheth2001} is less steep as a function of mass resulting from the differences of their underlying work to ours: First, they use the linear-shear model instead of the hybrid model, leading to a larger $\delta_\cc$ for all redshifts as shown in Fig.~\ref{fig:shearDeltas}. Second, the collapse of each axis is stopped using the artificial condition $a_i=0.177a$, leading to $\delta_\cc=1.686$ as a reference value at high mass. Thus, the fitting formula provides a larger value for large masses, while our approach leads to a $\delta_\cc=1.583$ for $M\rightarrow\infty$ due to the virialisation condition that we apply. Third, for a given initial overdensity $\delta$, \citet{Sheth2001} use the most probable value $e_\mathrm{mp}=(\sigma/\delta)/\sqrt{5}$ instead of the expectation value $\langle e\rangle$ given by Eq.~\eqref{eq:expectation}, which leads to initial ellipticities that are slightly too low so that the asymptotic limit for high redshifts is reached earlier than in our case.

The right panel of Fig.~\ref{fig:Sheth} shows a similar behaviour for $\delta_\cc$ as a function of redshift. The differences between the fitting formula and the result of our work again occur due to the different virialisation time and condition, and the difference between the most probable and the expectation value of $e$.

For applications, fitting formulae for both $\delta_\cc$ and $\Delta_\vv$ may be useful. We provide expressions here which are inspired by Eq.~\eqref{eq:fitting} of \citet{Sheth2001}. For $\delta_\cc(M,z)$ we suggest
\begin{equation}
\label{eq:fitDeltaC}
\delta_\cc(M,z)=\delta_\mathrm{c,sph}(z_\vv)\left\{1+b\left[\frac{\sigma^2(M,z_\vv)}{\delta_\mathrm{c,sph}^2(z_\vv)}\right]^c\right\}\,,
\end{equation}
where $b=0.6536$, $c=0.6387$, and both $\delta_\mathrm{c,sph}$ and $\sigma^2$ are cosmology-dependent quantities. Note again that the redshift of virialisation, $z_\vv$, has to be chosen as reference for $z$ for both $\delta_\mathrm{c,sph}$ and $\sigma^2$. For a spatially-flat $\Lambda$CDM model with matter density parameter in the range $\Omega_\mm\in[0.2,0.4]$, $M\in[10^{11},10^{15}]\ M_\odot\ h^{-1}$, and $z\in[0,10]$, the maximal error is $\sim$1.8\% with a mean error of $\sim$0.4\%.

A similar functional dependence can be found for $\Delta_\vv(M,z)$. Only a small correction term has to be added to arrive at a satisfactory accuracy. We find
\begin{equation}
 \label{eq:fitDeltaV}
\Delta_\vv(M,z)=\Delta_\mathrm{v,sph}(z_\vv)\left[a+b\,\sigma^{2c}(M,z_\vv)+d\,(1+z_\vv)^{2/5}\log^{9/4}(M)\right]
\end{equation}
with $a=0.3819$, $b=0.5379$, $c=0.7589$, and $d=3.456\times10^{-4}$. In the same range as above, the maximal error is $\sim$5\% with a mean error of $\sim$1\%.

\subsection{Influence of initial ellipticity and prolaticity}
\label{subsec:ep}

\begin{figure}
\resizebox{\hsize}{!}{\includegraphics{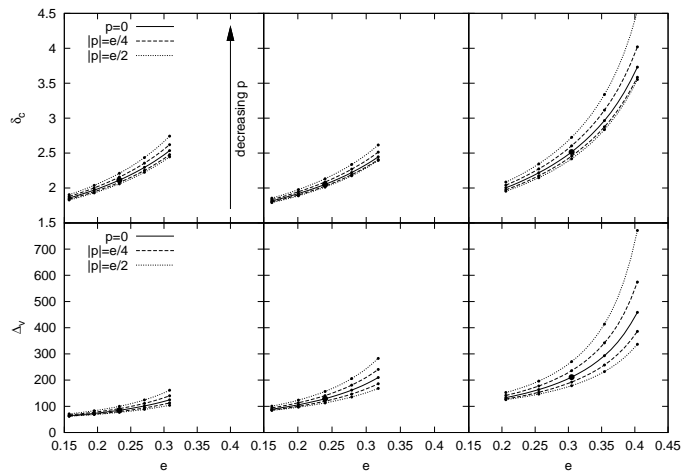}}
\caption{Influence of the initial ellipticity $e$ and prolaticity $p$ on the parameters $\delta_\cc$ (\emph{top panels}) and $\Delta_\vv$ (\emph{bottom panels}) for three different cosmological models. \emph{Left panels:} $\Lambda$CDM. \emph{Central panels:} OCDM. \emph{Right panels:} EdS. The large circle in each figure indicates the value of the respective parameter for $e=\langle e\rangle$ and $p=0$. The small circles indicate values for a combination of $e\in[\langle e\rangle \pm \sigma_e/2,\ \langle e\rangle \pm \sigma_e]$ and $p\in[\pm e/4,\ \pm e/2]$ (see Eqs.~\ref{eq:expectation}, \ref{eq:variances}). For a fixed $e$ both $\delta_\cc$ and $\Delta_\vv$ grow for decreasing $p$.}
\label{fig:influenceEP}
\end{figure}

In Fig.~\ref{fig:influenceEP} we plot both $\delta_\cc$ and $\Delta_\vv$ as a function of the initial ellipticity $e$ and prolaticity $p$ centered around their expectation values given by Eq.~\eqref{eq:expectation} for three different cosmologies. For increasing $e$ and decreasing $p$, both parameters grow qualitatively in the same way as already reported by \citet{Sheth2001} (cf.~their Fig.~1). Quantitative deviations arise from the differences in the applied algorithm as discussed in Sect.~\ref{subsec:parameters}. For a given mass and virialisation redshift, the initial overdensity for the EdS universe is larger compared to both $\Lambda$CDM and OCDM due to a shorter physical time interval that corresponds to the same redshift interval, resulting in a larger $\langle e \rangle$ and $\sigma_e$, but also in larger curvatures of $\delta_\cc$ and $\Delta_\vv$ with respect to $e$ and $p$. These are the sources of the larger error in the approximation $\langle\xi\rangle\approx\xi(\langle e \rangle,\langle p\rangle)$ discussed in Sect.~\ref{subsec:epIni}. Since the redshift-time relation is not very different between $\Lambda$CDM and OCDM, the dependences of $\delta_\cc$ and $\Delta_\vv$ on $e$ and $p$ are comparable.

\subsection{Mass function}
\label{subsec:massFunction}

Using Eq.~\eqref{eq:fitDeltaC}, we are able to construct the mass function of dark-matter haloes using the extended Press-Schechter formalism developed by \citet{Bond1991} and \citet{Lacey1993}, which is based on the first-upcrossing distribution of the density contrast $\delta$ as a function of the ``time variable'' $S\equiv\sigma^2(M)$. We shall proceed similarly as \citet{Sheth1999,Sheth2002} and define the scaled variable $\nu\equiv\delta_\mathrm{c,sph}^2/S$ to derive the mass function for our standard $\Lambda$CDM cosmology.

As \citet{Sheth2002} pointed out, expressing the first-upcrossing distribution $f$ as a function of $\nu$ has the advantage that it is only necessary to calculate $f(\nu)$ for a barrier of height $B(\nu,z)$ at one arbitrary redshift to infer the mass function $n(M)$ at any other redshift by a simple rescaling. For a given first-upcrossing distribution $f(\nu)$, the differential mass function can be calculated using the relation
\begin{equation}
\label{eq:defineMassFunction}
n(M)=\frac{\rho_\mathrm{b}}{M^2}\frac{\dd\ln\nu}{\dd\ln M}\nu f(\nu)\,,
\end{equation}
where $\rho_\mathrm{b}$ is the background density of the Universe.

\begin{figure}[t]
\resizebox{\hsize}{!}{\includegraphics{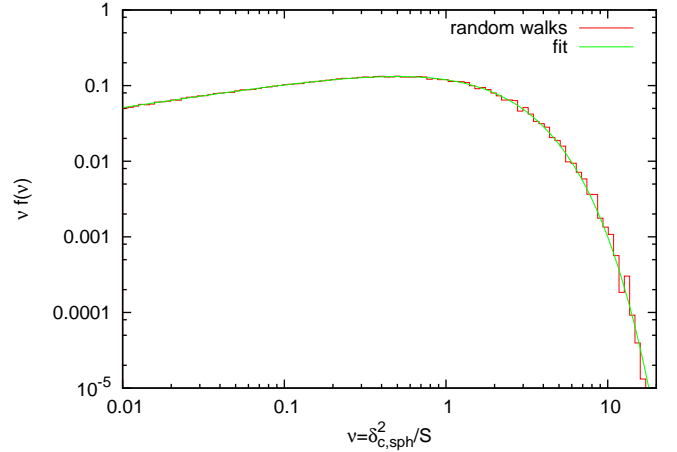}}
\caption{Comparison between the first-upcrossing distribution of the moving barrier, Eq.~\eqref{eq:barrier}, inferred from an ensemble of one million random walks, and the fitting formula, Eq.~\eqref{eq:fitFirstUp}.}
\label{fig:firstUpDist}
\end{figure}

First, we want to find an accurate fit to the first-upcrossing distribution of a moving barrier which is given by the mass-dependent linear overdensity parameter of the ellipsoidal collapse,
\begin{equation}
\label{eq:barrier}
B(\nu)=\delta_\mathrm{c,sph}\left(1+0.6536\,\nu^{-0.6387}\right)
\end{equation}
(see Eq.~\ref{eq:fitDeltaC}). The parameter $\delta_\mathrm{c,sph}$ is evaluated at $z_\vv=0$. We ran one million random walks and recorded the first-upcrossing values for $\nu\in[0.01,20]$ in 100 equidistant bins in logarithmic space. The resulting distribution $\nu f(\nu)$ is nicely expressed by the function
\begin{equation}
\label{eq:fitFirstUp}
\nu f(\nu)=A\left[1+(a\nu)^{-p}\right]\sqrt{\frac{a\nu}{2\pi}}\exp\left(-\frac{a\nu}{2}
\frac{B(\nu)}{\delta_\mathrm{c,sph}}\right)\,,
\end{equation}
Thus, our suggested fitting formula is a mixture of the functional forms proposed by \citet{Sheth1999} and \citet{Sheth2002}. The remaining best-fit parameters are $A=0.357$, $p=0.212$ and $a=1.171$. The result is shown in Fig.~\ref{fig:firstUpDist}.

\begin{figure}[t]
\resizebox{\hsize}{!}{\includegraphics{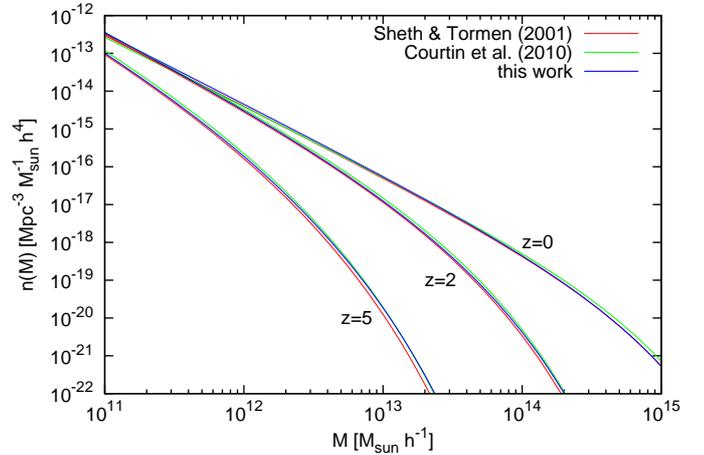}}
\caption{Comparison of the \citeauthor{Sheth1999} and the \citeauthor{Courtin2010} mass functions with the mass function derived from the rescaled upcrossing distribution, Eq.~\eqref{eq:fitFirstUp}, based on our treatment of the ellipsoidal-collapse dynamics.}
\label{fig:massFunctions}
\end{figure}

Second, to find a viable mass function from the first-upcrossing distribution, we proceed as \citet{Sheth1999} and \citet{Sheth2001}, normalise $f(\nu)$ to unity and rescale the variable $a$ such that we are in agreement with the standard \citeauthor{Sheth1999} mass function and a mass function based on $N$-body simulations proposed by \citet{Courtin2010}. The latter is based on a first-upcrossing distribution that has the same functional form as that proposed by \citet{Sheth1999}, but slightly different best-fit parameters,
\begin{equation}
\label{eq:Courtin}
\nu f_\mathrm{Courtin}(\nu)=\tilde{A}\left[1+(\tilde{a}\nu)^{-\tilde{p}}\right]\sqrt{\frac{\tilde{a}\nu}{2\pi}}\exp\left(-\frac{\tilde{a}\nu}{2}\right)\,,
\end{equation}
with $\tilde{A}=0.348$, $\tilde{a}=0.695$, and $\tilde{p}=0.1$. Note that in their definition of $\nu$, the linear density contrast $\delta_\mathrm{c,sph}$ has to be taken at \emph{collapse}. Normalising the first-upcrossing distribution based on the moving barrier of our ellipsoidal-collapse model to unity yields a rescaled parameter $A\rightarrow A'=1.364\,A$. We compare the resulting mass function with those by \citet{Sheth1999} and \citet{Courtin2010} for three different redshifts in Fig.~\ref{fig:massFunctions}. The parameter $a$ was rescaled by $a\rightarrow a'=0.625\,a$. Deviations from the \citeauthor{Sheth1999} mass function at high masses occur at large redshifts which is compatible with the \citeauthor{Courtin2010} mass function. Overall, our proposed mass function lies in between these two, suggesting that differences between $N$-body simulations and the \citeauthor{Sheth1999} mass function might be due to an imprecise treatment of the ellipsoidal-collapse dynamics.

\section{Conclusions}
\label{sec:conclusions}

We have reconsidered the collapse of a homogeneous, triaxial ellipsoid in an expanding background universe and extended the treatment of \citet{Bond1996} in two ways:

\begin{itemize}

\item We have introduced a physically motivated criterion for the onset of virialisation along each principal axis of a collapsing ellipsoid. We derive this criterion from the tensor virial theorem, demanding that the inertial tensor stabilises and its second time derivative vanishes. This approach is a generalisation of the usual virialisation condition for the spherical-collapse scenario and thus fully consistent with it. It replaces the conventional requirement that virialisation is assumed along an axis $i$ when the dimension-less semi-major axis $a_i$ of the ellipsoid along that axis reaches $a_i(a)=0.177a$.

\item We have introduced a hybrid model for the influence of the external gravitational shear field acting on the ellipsoid, supplementing the two different models of \citet{Bond1996}. The principal axes of the shear field evolve with those of the halo before it turns around and decouples from the background expansion, and then continues to evolve linearly while the halo evolves non-linearly and collapses. We have shown that the differences between the hybrid and the non-linear model are relatively small.

\end{itemize}

For a given initial ellipticity $e$ and prolaticity $p$, and for a specified virialisation redshift $z_\vv$, the ellipsoidal-collapse model then gives a unique answer for the linear density contrast $\delta_\cc$ at virialisation, as well as for the overdensity $\Delta_\vv$ at that time. The probability distribution for $e$ and $p$, conditional on the density contrast $\delta$, is determined by the probability distribution of the eigenvalues of the Zel'dovich deformation tensor, as shown by \citet{Sheth2001}. It is characterised by variance of the matter-density fluctuations on a scale fixed by the halo mass required. We have shown that the marginalisation over $e$ and $p$ can be simplified by evaluating the ellipsoidal collapse at their mean values $\langle e\rangle$ and $\langle p\rangle=0$.

Our main results are as follows:

\begin{itemize}

\item The collapse parameters $\delta_\cc$ and $\Delta_\vv$ depend only weakly on the model for the external gravitational shear. The hybrid model and the non-linear model by \citet{Bond1996} give approximately the same results.

\item When supplied with our virialisation condition derived from the tensor virial theorem, the ellipsoidal-collapse model returns values $\delta_\cc$ and $\Delta_\vv$ that differ substantially from those obtained with the spherical-collapse model. Depending on halo mass and redshift, deviations of order $(20\ldots50)\%$ are common. After marginalisation over $e$ and $p$, $\delta_\cc$ and $\Delta_\vv$ increase with decreasing halo mass and with decreasing virialisation redshift.

\item Both parameters increase with increasing initial ellipticity $e$ and decrease with increasing prolaticity $p$, as already suggested by \citet{Sheth2001}.

\item Our results for $\delta_\cc$ qualitatively confirm the dependence on halo mass and virialisation redshift given by a fitting formula by \citet{Sheth2001}. Deviations in particular at low mass and high redshift occur due to the differences in the virialisation condition, the model for the external shear, and the marginalisation over $e$.

\item The mass function based on our refined treatment of the ellipsoidal-collapse model is in good agreement with those proposed by \citet{Sheth1999} for low redshifts and \citet{Courtin2010} for high redshifts. This suggests that differences between the \citeauthor{Sheth1999} mass function and results from $N$-body simulations at large redshifts may occur due to an imprecise treatment of the ellipsoidal-collapse dynamics.

\end{itemize}

We now plan to introduce the ellipsoidal-collapse model into the approach presented by \citet{Angrick2009} to improve the prediction of the X-ray temperature function for galaxy clusters based on the gravitational potential rather than on the density contrast.

\acknowledgements{CA wants to thank Francesco Pace for fruitful discussions, and the Deutsche Forschungsgemeinschaft for financial support under grant number BA 1369/12-1, the Heidelberg Graduate School of Fundamental Physics, and the IMPRS for Astronomy \& Cosmic Physics at the University of Heidelberg.}

\normalsize

\appendix
\section{Comparison to a previous study}
\label{ap:monaco}

Here we compare our results for the evolution equations of the collapsing ellipsoid with those presented by \citet{Monaco1997} for a flat $\Lambda$CDM and an OCDM model.

Starting from Eq.~\eqref{eq:basicEvolutionA}, we can replace $E(a)$ and $E'(a)$ for a flat $\Lambda$CDM model using
\begin{equation}
\label{eq:expansionLCDM}
E(a)=\sqrt{\Omega_\mm a^{-3}+(1-\Omega_\mm)}\,,\quad E'(a)=-\frac{3\Omega_\mm a^{-4}}{2E(a)}
\end{equation}
since $\Omega_\Lambda=1-\Omega_\mm$ and the curvature parameter $\Omega_\mathrm{k}=0$. This gives
\begin{equation}
 \label{eq:evolutionLCDM}
\frac{\dd^2 a_i}{\dd a^2}-\frac{1-2(\Omega_\mm^{-1}-1)a^3}{2a\left[1+(\Omega_\mm^{-1}-1)a^3\right]}\frac{\dd a_i}{\dd a}+\frac{3 C_i-2a^3(\Omega_\mm^{-1}-1)}{2a^2\left[1+(\Omega_\mm^{-1}-1)a^3\right]}a_i=0\,.
\end{equation}
This equation differs from Eq.~(B11) of \citet{Monaco1997} in the second and third term: a factor $a^2$ in the denominator of both terms was omitted. Additionally, the vacuum term $\propto \rho_\Lambda$ as well as a factor $3$ were not included in the nominator of the third term.

For an OCDM model, we have
\begin{align}
\label{eq:OCDMe}
E(a)&=\sqrt{\Omega_\mm a^{-3}+(1-\Omega_\mm)a^{-2}}\,, \\
\label{eq:OCDMePrime}
E'(a)&=-\frac{3\Omega_\mm a^{-4}+2(1-\Omega_\mm) a^{-3}}{2E(a)}
\end{align}
since $\Omega_\Lambda=0$ and $\Omega_\mathrm{k}=(1-\Omega_\mm)a^{-2}$. Inserting this again into Eq.~\eqref{eq:basicEvolutionA} yields
\begin{equation}
 \label{eq:evolutionOCDM}
\begin{split}
\frac{\dd^2 a_i}{\dd a^2}-\left\{2a\left[1+(\Omega_\mm^{-1}-1)a\right]\right\}^{-1}\frac{\dd a_i}{\dd a}&\\
+3\left\{2a^2\left[1+(\Omega_\mm^{-1}-1)a\right]\right\}^{-1}C_i a_i&=0\,.
\end{split}
\end{equation}
Equation~(B12) of \citet{Monaco1997} is again slightly different: The factor $3$ in the last term was omitted.

There is one last difference concerning the initial conditions: Comparing Eq.~(B17) of \citet{Monaco1997} with Eq.~\eqref{eq:initAxes} of this paper, one can find an additional factor $a_0$ in front of $\lambda_i(a_0)$ which should be dropped.

\bibliography{14147.bib}

\end{document}